\documentclass[]{scrartcl}
\usepackage{graphics}
\usepackage[english]{babel}
\usepackage{amsmath} 
\usepackage{amsfonts}
\usepackage{latexsym}
\usepackage{cuted}
\usepackage{graphicx}
\usepackage{graphics}
\usepackage{float}
\usepackage{breqn}
\usepackage{amssymb}
\usepackage{authblk}
\usepackage{makeidx}
\usepackage{dblfloatfix}
\usepackage[section]{placeins}

\PassOptionsToPackage{hyphens}{url}
\usepackage{hyperref}
\usepackage{url}
\begin{document}

\title{Reply to ``A comment on ``A test of general relativity using the LARES
and LAGEOS satellites and a GRACE Earth gravity model,
by I. Ciufolini et al.'' ''}
\author[1,2]{Ignazio Ciufolini\thanks{ignazio.ciufolini@unisalento.it}}
\author[3]{Erricos C. Pavlis}
\author[4]{John Ries}
\author[5]{Richard Matzner}
\author[6]{Rolf Koenig}
\author[7]{Antonio Paolozzi}
\author[7]{Giampiero Sindoni}
\author[8]{Vahe Gurzadyan}
\author[9]{ Roger Penrose}
\author[2,7]{Claudio Paris}

\affil[1]{\footnotesize Dip. Ingegneria dell'Innovazione, Universit\`a del Salento, Lecce, Italy}
\affil[2]{Museo della fisica e Centro studi e ricerche Enrico Fermi, Rome, Italy}

\affil[3]{Joint Center for Earth Systems Technology (JCET), University of Maryland, Baltimore County, USA}

\affil[4]{Center for Space Research, University of Texas at Austin, Austin, USA}
\affil[5]{Theory Center, University of Texas at Austin, Austin, USA}

\affil[6]{Helmholtz Centre Potsdam, GFZ German Research Centre for Geosciences, Potsdam, Germany}
\affil[7]{Scuola di Ingegneria Aerospaziale, Sapienza Universit\`a di Roma, Italy}
\affil[8]{Center for Cosmology and Astrophysics, Alikhanian National Laboratory and Yerevan State University, Yerevan, Armenia}

\affil[9]{Mathematical Institute, University of Oxford, Oxford, UK}

\renewcommand\Authands{ and }

\date{}
\maketitle

\abstract{In 2016, we published ``A test of general relativity using the LARES and LAGEOS satellites and a GRACE Earth gravity model.
Measurement of Earth’s dragging of inertial frames \cite{ciuetal16}",  a measurement of frame-dragging, a fundamental prediction of Einstein's theory of General Relativity, using the laser-ranged satellites LARES, LAGEOS and LAGEOS 2. The formal error, or precision, of our test was about 0.2\% of frame-dragging, whereas the systematic error was estimated to be about 5\%.
In the 2017 paper ``A comment on ``A test of general relativity using the LARES and LAGEOS satellites and a GRACE Earth gravity model by I. Ciufolini et al.'' '' by L. Iorio \cite{ior} (called I2017 in the following), it was incorrectly claimed that, when comparing different Earth gravity field models, the systematic error in our test due to the Earth's even zonal harmonics of degree 6, 8, 10 could be as large as 15\%, 6\% and 36\%, respectively. Furthermore, I2017 contains other, also incorrect, claims about the number of necessary significant decimal digits of the coefficients used in our test (claimed to be nine), in order to eliminate the largest uncertainties in the even zonals of degree 2 and 4, and about the non-repeatability of our test. Here we analyze and rebut those claims in I2017.

\section{Introduction}
\label{intro}

The dragging of inertial frames, or frame-dragging, is a fundamental and intriguing prediction of Einstein's theory of General Relativity. It has a key role in a number of astrophysical phenomena, including the orientation of jets from active galactic nuclei and quasars and the emission of gravitational waves from colliding black holes (\cite{tho,gw}). In General Relativity, the angular momentum of a central body causes a secular shift of the nodes of a satellite (the intersections of its orbit with the equatorial plane of the central body), and of its periastron (the closest point of its orbit to the central body) around that central body. This is called Lense-Thirring effect (\cite{lent}). In a number of papers \cite{ciu84,ciu86,ciu89,csr89,asi89,rie89,pet,ciu96}, we described how, by combining the orbital elements of a number of satellites with suitable coefficients, it would be possible to test frame-dragging and the Lense-Thirring effect with an accuracy depending on the number of the satellites' orbital observables used in the analysis and on their accuracy. The technique is described in detail in \cite{ciu96}; here we simply note that the major systematic errors arise from errors in the Earth's even zonal harmonics (the Earth's deviations from spherical symmetry which are both symmetrical with respect to the Earth's equatorial plane and to its symmetry axis.) In particular the largest source of systematic error is due to the largest deviation of the Earth from spherical symmetry, its oblateness, described by the even zonal harmonic of degree two, the Earth's quadrupole moment. Indeed each even zonal harmonic generates a {\\itshape classical} (i.e. not General Relativistic) shift of the node of a satellite and these shifts are dominated by the lowest degree even zonal harmonics and especially by the Earth's quadrupole moment. An idea \cite{ciu86} was to use two laser-ranged satellites with supplementary inclinations to eliminate the error due to the uncertainties of all the even zonal harmonics (this technique will be achieved by the forthcoming LARES 2, Laser Relativity Satellite 2, of ASI - the Italian Space Agency). Another idea was then to use $n$ observables, and in particular the $n$ nodes of $n$ satellites to both measure the Lense-Thirring effect and to eliminate the uncertainties due to the largest $n - 1$ even zonal harmonics: ``Another solution would be to orbit several high-altitude, laser-ranged satellites, similar to LAGEOS, to measure $J_2, J_4, J_6$ etc., and one satellite to measure $\dot{\Omega}^{Lense-Thirring}$ ''  (p. 3102 of \cite{ciu89}).

A number of tests \cite{ciupav,ciupavper,ciuetal10,ciuetal16} with ever increasing accuracy was then carried out using this last technique, first using the two satellites LAGEOS (1976) of NASA and LAGEOS 2 of ASI and NASA (1992 \cite{coen}), both originally dedicated to space geodesy, and then including  LARES (Laser Relativity Satellite), launched in 2012 by ASI, dedicated to relativity and space geodesy. In 2016, we published  \cite{ciuetal16} a test of the Lense-Thirring effect using about 3.5 years of data of LARES, LAGEOS and LAGEOS 2. This test used their three nodal observables to eliminate the error due to the first two largest even zonal harmonics, i.e., the Earth quadrupole moment $J_2$, of degree two, and the even zonal of degree four $J_4$, and to test the Lense-Thirring effect. The formal error, or precision, of our test was about 0.2\% of frame-dragging, whereas the systematic error was estimated to be about 5\%. This systematic error was mainly due to the even zonal harmonics of degree strictly higher than four and was calculated by using the calibrated errors (i.e. including the systematic errors) of the Earth gravity model GGM05S \cite{GGM05S,reisEtAl2016} which we use to specify moderately low angular components of the Earth's gravity field. (In our analysis the Earth model GGM05S provided the even zonal harmonics of degree $2n = 6, 8, . . . , 90$. The large - degree harmonics have very little effect on the results.) GGM05S is a state-of-the-art determination of the Earth gravity field, obtained using the space mission GRACE (Gravity Recovery and Climate Experiment), launched in 2002  \cite{gra}. GRACE determined the Earth gravity field and its variations using two spacecraft in polar orbit at an altitude of about 400 kilometers. The pair extracted variations in the gravitational field by accurate  ranging to each other.

A recent paper ``A comment on ``A test of general relativity using the LARES and LAGEOS satellites and a GRACE Earth gravity model by I. Ciufolini et al.'' ''  by L. Iorio \cite{ior} (called I2017 in the following), claims, based on a comparison among different Earth gravity field models, that the systematic errors of our 2016 test, due to the Earth's even zonal harmonics of degree 6, 8 and 10, can be as large as 15\%, 6\% and 36\%, respectively. We show below Iorio is incorrect in these claimed results. In fact, I2017 mentions the Earth gravity model we use (GGM05S) only three times: once in the abstract, once in section 2.2 and once in the comment: ``It can be noted that Eq. (31) yields a realistic uncertainty for $C_{6,0}$ very close to the simple difference $C_{6,0}$ between the estimated coefficients of ITU\textunderscore GRACE16 and GGM05S''. ITU\textunderscore GRACE16 is another Earth model.  Eq. (31) in I2017 calculates a coefficient differencing ITU\textunderscore GRACE16 and yet another Earth model: GOCO05S. So, on its face, I2017 says nothing directly about the accuracy of GGM05S but uses an arbitrary selection of models to infer the accuracy of the degree 6, 8, and 10 zonal harmonics.

In section 2.1 we show that the systematic errors reported in the paper  I2017  are incorrect by some substantial factors. In section 2.2 we show that, with regard to the accuracy of the lowest even zonal harmonics, at least two of the Earth gravity models used in  I2017, i.e., JYY\textunderscore GOCE04S and ITU\textunderscore GRACE16 are not comparable in accuracy with the Earth gravity model GMM05S we use, obtained with GRACE. In particular the lowest harmonics of the model JYY\textunderscore GOCE04S, obtained using data from the space mission Gravity Field and Steady-State Ocean Circulation Explorer (GOCE) \cite{GOCE} $only$, cannot be compared with the accuracy of the lowest harmonics of the GRACE and Satellite Laser Ranging (SLR) model GGM05S. GOCE was designed to generate gravity field models with increased accuracy for the higher degree harmonics of the Earth's gravitational field but is not comparable in accuracy to GRACE (about an order of magnitude worse) for the lowest harmonics = the ones that dominate the errors in the Lense-Thirring analysis.

I2017 contains other, incorrect, claims about the number of  significant decimal digits of the coefficients used in our test (claimed to be nine), necessary to eliminate the largest uncertainties in the even zonal of degree 2 and 4, and about the non-repeatability of our test, and other minor claims. In section 3, we show that the claim of I2017 that nine significant decimal digits in the coefficients are necessary for the cancellation of the error due to $J_2$ and $J_4$ is not correct and in fact, for a 1\% test of frame-dragging, we only need two or three significant decimal digits. Finally in Section 3.1, we address the claim of  I2017 about the non-repeatability of our test of frame-dragging, and other minor claims.

\section{Erroneous claims of the errors induced by the gravity field uncertainties}

In I2017 the even zonal harmonics $\bar{C}_{6,0}$, $\bar{C}_{8,0}$ and $\bar{C}_{10,0}$ of the gravity field models ITU\textunderscore GRACE16,\\ ITSG\textunderscore Grace2014S, GOCO05S and JYY\textunderscore GOCE04S are compared. (The $\bar{C}_{2n,0}$ are related by a normalization to the even zonal harmonics $J_{2n}$. The explicit relation is given in section 2.1 below.) The difference between each normalized even zonal harmonic, $\bar{C}_{6,0}$, $\bar{C}_{8,0}$ and $\bar{C}_{10,0}$, of each pair of these gravity models is then calculated (see tables 3, 5 and 7 of I2017), and these differences are then propagated into the combination of the nodes of LAGEOS, LAGEOS 2 and LARES to produce a claimed percent error in the measurement of the frame-dragging of their nodes, i.e., of the Lense-Thirring effect (see tables 4, 6, 8 and 9 of I2017).

However, the findings of I2017 are affected by erroneous claims, both numerical and conceptual, as we now show.

\subsection{Numerical miscalculations in I2017}

In I2017 Iorio claims that the errors induced in the test of frame-dragging by the differences in the coefficients $\bar{C}_{6,0}$, $\bar{C}_{8,0}$ and $\bar{C}_{10,0}$ of the four above models are quite large and, for example, the errors induced by the differences in $\bar{C}_{6,0}$ may be as large as 15\% of frame-dragging. Similar claims are made for $\bar{C}_{8,0}$ and $\bar{C}_{10,0}$.

Let us concentrate on the errors due to $\bar{C}_{6,0}$. We use the treatment of the standard text of space geodesy by Kaula \cite{kau}; we have also checked the results with the orbital estimator GEODYN. We find that the secular rate of the node of a satellite due to $\bar{C}_{6,0}$ can be easily calculated as follows.

The Lagrange equation for the rate of change of the node $\Omega$ of a satellite as a function of a disturbing function $R$ is \cite{kau,bert}:

\begin{equation}
\frac{d\Omega}{dt}=\frac{1}{na^{2}(1-e^{2})^{1/2}\sin i}\frac{\partial F}{\partial i}
\end{equation}

Where the force function $F$ is given by $F = \frac{G \, M_\oplus}{2a}+R$, $G$ is the gravitational constant, $M_\oplus$ is the Earth mass, and $n, a, e$ and $i$ are
respectively mean motion, semimajor axis, orbital eccentricity and inclination of an Earth satellite.

The disturbing function $R$ depends on the Earth potential $V$ (not including the central term).
The Earth’s potential $V$, real solution of the Laplace equation, can be written \cite{kau}:

\begin{equation*}
V=\sum _{l=0}^{\infty} \sum _{m=0}^{n} \frac{1}{r^{l+1}}P_{lm}(\sin\phi )[C_{lm}\cos m\lambda + S_{lm}\sin m\lambda]
\end{equation*}

where  $P_{lm}(\sin\phi)$ are the Legendre associated functions, $r$, $\phi$ and $\lambda$ are respectively radial coordinate, latitude and longitude measured eastward, $l$ and $m$ are degree and order of the spherical harmonic, and $C_{lm}$ and $S_{lm}$ are respectively the cosine and sine coefficients of the spherical harmonic potential term.
The term $V_{lm}$ of the Earth potential of degree 6 and order 0 due to the even zonal harmonic $\bar{C}_{6,0}$ can be written \cite{kau,bert,geo}:

\begin{equation}
V_{60}=\frac{G \, M_{\oplus} \, R_{\oplus}^{6}}{a^{6+1}} \sum_{p=0}^{6}F_{60p}(i) \sum_{q=-\infty}^{\infty} G_{6pq}(e) S_{60pq}(\omega , M, \Omega)
\end{equation}

where:

\begin{dmath}
S_{60pq}= \sqrt{13}\, \bar{C}_{6,0} \cos [(6 - 2p) \omega + (6 - 2p + q) M]
\end{dmath}

\noindent and $R_{\oplus}$, $\omega$ and $M$ are respectively Earth radius, satellite' argument of perigee and mean anomaly.
$\bar{C}_{6,0}$ is the normalized even zonal harmonic coefficient of degree 6 and order 0.
The normalized even zonal harmonic coefficients, $\bar{C}_{2n, \, 0}$, the ones usually provided in the Earth gravity field models, are related to the denormalized coefficients $C_{2n, \,0}$ by the simple relation: $C_{2n, \,0} \equiv  \sqrt{4n + 1} \bar{C}_{2n, \, 0})$.
 For example $C_{20} = -1.8264 \cdot 10^{-3}$ and $\bar{C}_{20} = -4.8417 \cdot 10^{-4}$, and
 $C_{6,0} = -5.40743 \cdot 10^{-7}$ and
 $\bar{C}_{6,0} = -1.49975 \cdot 10^{-7}$  i.e.,
  for the degree six even zonal harmonic: $C_{6,0} \equiv \sqrt{13} \bar{C}_{6,0}$,
   (the non-normalized even zonal harmonic coefficients, usually written with the notation $J_{2n}$
   are equal to the $C_{2n, \, 0}$ coefficients with a minus sign, e.g., the quadrupole coefficient
    $J_{2}$ is $J_2 = 1.8264 \cdot 10^{-3}$).

By considering the secular rate only of the nodes of a satellite due to the even zonal harmonic of degree 6, $\bar{C}_{6,0}$, we have then:
\begin{equation}
V_{60} = \frac{G M_{\oplus} \sqrt{13}\bar{C}_{6,0} }{a}\left( \frac{R_{\oplus}}{a} \right)^{6} F_{603}(i) G_{630}(e)
\end{equation}
The functions $F_{603}(i)$ and  $G_{630}(e)$ can be easily calculated using the recursive formulae of Kaula and are given by $F_{603}=-\frac{5}{16}+\frac{105(\sin)^{2} i}{32}-\frac{945(\sin)^{4} i}{128}+\frac{1155(\sin)^6 i}{256}$ and\\ $G_{630}=\frac{1+5e^{2}+\frac{15e^{4}}{8}}{(1-e^{2})^{11/2}}$:

Finally inserting $F_{603}(i)$ and  $G_{630}(e)$ in Eqs. 1 and 4, we have the secular nodal rate due to $\bar{C}_{6,0}$:

\begin{dmath}
\frac{d\Omega_{6,0}}{dt} = \frac{105 (1+5e^{2}+\frac{15 e^{4}}{8}) n R^{6}\sqrt{13}\, \bar{C}_{6,0}}{16a^{6}(1-e^{2})^{6}} \cdot    \cos i (1-\frac{9(\sin)^{2} i}{2}+\frac{33(\sin)^{4} i}{8}) 
\end{dmath}

By inserting in the nodal rate the orbital parameters, semimajor axis, $a$, eccentricity, $e$, and inclination, $i$, of the three satellites:
$ a_{LARES} \cong$ 7820 km, $ e_{LARES} \cong$ 0.0008, and $ i_{LARES} \cong$ 69.5$^{\circ}$;
$ a_{LAGEOS} \cong$ 12,270 km, $ e_{LAGEOS} \cong$ 0.0045, and $ i_{LAGEOS} \cong$ 109.84$^{\circ}$, and
$ a_{LAGEOS \, 2} \cong$ 12,163 km, $ e_{LAGEOS \, 2} \cong$ 0.0135, and $ i_{LAGEOS \, 2} \cong$ 52.64$^{\circ}$; we have:

\begin{eqnarray*}
\frac{d\Omega_{LAGEOS}}{dt} = -1.18019 \cdot 10^{11} \cdot \bar{C}_{6,0}\; mas/yr\\
\frac{d\Omega_{LAGEOS \, 2}}{dt} = -1.78652 \cdot 10^{11} \cdot \bar{C}_{6,0}\; mas/yr\\
\frac{d\Omega_{LARES}}{dt} = 3.27064 \cdot 10^{12} \cdot \bar{C}_{6,0}\; mas/yr\\
\end{eqnarray*}

Where mas stands for milliarcsec. Combining the nodal rates of LAGEOS, LAGEOS 2 and LARES due to $\bar{C}_{6,0}$ using the formula to eliminate the $\bar{C}_{2,0}$ and $\bar{C}_{4,0}$ contributions to the combined nodal rates (see formula (9) of section 3 below), we have:\\
\begin{strip}
\begin{dmath}
{ \Omega}^{6,0}_{LAGEOS} + c_1 { \Omega}^{6,0}_{LAGEOS 2} +
c_2 { \Omega\/}^{6,0}_{LARES} = (-1.18019 \cdot 10^{11} - c_1 \cdot 1.78652 \cdot 10^{11} + c_2 \cdot 3.27064 \cdot 10^{12}) \cdot \bar{C}_{6,0}\; mas/yr =
5.91029 \cdot 10^{10} \cdot \bar{C}_{6,0} \; mas/yr
\end{dmath}
\end{strip}

\noindent where $c_1 = 0.345$ and $c_2 = 0.073$.

Finally, the largest $C_{6,0}$ difference in Iorios's Table 3 (I2017) is GOCO05S - ITU\textunderscore GRACE16:  $\Delta \bar{C}_{6,0} = 3.197 \times 10^{-11}$ in magnitude. Using this difference we get the error in the combined nodal rates of LAGEOS, LAGEOS 2 and LARES due to the difference between the $\bar{C}_{6,0}$ coefficients of GOCO05S and ITU\textunderscore GRACE16, that is 1.89 mas/year.

Since the combined frame-dragging effect is about\\ ${ \Omega}^{Lense-Thirring}_{combination} = 30.657 + c_1 \cdot 31.481 + c_2 \cdot 118.421 \; mas/yr =$\\
$ 50.16 \; mas/yr$, the final relative percent error  is just:

\begin{equation}
\frac{1.89 \; mas/yr}  {50.16  \; mas/yr} = 3.75 \% \; { \Omega}^{Lense-Thirring}_{combination},
\end{equation}

\noindent an error about four times smaller than 15\% as erroneously claimed in I2017, and within our 5\% estimated systematic error.  Other entries in I2017 Table 3 are smaller (or much smaller) than GOCO05S - ITU\textunderscore GRACE16; the effect on the error is linear in the differences, so this result bounds the Lense-Thyrring error estimate derived from $C_{6,0}$ differences.

Similar calculational/numerical errors affect the other values listed in tables 4, 6, 8 and 9 of I2017. To continue our analysis of the difference, we find the percentage uncertainty arising from the difference in $C_{8,0}$ to be $3\times 10^{-3}\%$, compared to $2\times 10^{-2}\%$  in I2017.  For the percentage uncertainty arising from the $C_{10,0}$ difference, we find, in agreement with I2017, $\approx 3\%$. Obviously, adding the uncertainties arising from $C_{6,0}$, $C_{8,0}$, and $C_{10,0}$ would lead to $\approx 6.75\%$ added in absolute value, and about $4.8\%$ added in quadrature. The discussion just above concerns models GOCO05S and ITU\textunderscore GRACE16. Neither is the model GGM05S that we use, but COCO05S is very similar to GGM05S, and has similar good low-multipole accuracy. ITU\textunderscore GRACE16 has much poorer low-multipole accuracy, and as we have just seen, this leads to estimated frame dragging uncertainty  in the $5\%$ to $7\%$ range arising from differencing $C_{6,0}$, $C_{8,0}$, and $C_{10,0}$ between GOCO05S and ITU\textunderscore GRACE16.

The strongest claim made in I2017 involves differences involving $C_{10,0}$ between  the model JYY\textunderscore GOCE04S and the other three models considered in I2017. The $C_{10,0}$ differences between the model JYY\textunderscore GOCE04S and the others considered in I2017 would lead to frame dragging uncertainties of order $30\%$. (However JYY\textunderscore GOCE04S is about an order of magnitude less accurate than state of the art models in the low multipoles; see Fig. 1. I2017's calculations are erroneous also here.  I2017's Table 8,  last column (JYY\textunderscore GOCE04S) should read $32\%$, $29\%$, $32\%$.) Once these and other computational errors in I2017 are corrected, these $\approx 30\%$ differences dominate Iorio's claims for large ``uncertainties". But reviewing I2017's Tables 6 and 8  most clearly shows that  model JYY\textunderscore GOCE04S is an outlier; the fault lies with JYY\textunderscore GOCE04S (see Fig. 1). I2017's claims based on this outlier are not credible.

It is worth mentioning that in the comparison of \\ ITU\textunderscore GRACE16 and GGM05S, the effective epoch of the zonals can be different, which is relevant if they have a linear time dependence (seasonal and tidal variations do not have a significant impact on the results). GGM05S was determined with GRACE data spanning April 2003 to May 2013 (making the effective epoch $\sim$ 2008), while ITU\textunderscore GRACE16 used GRACE data from April 2009 to October 2013 (making the effective epoch $\sim$ 2011). Taking into account the linear drift (as determined from the full GRACE time series currently available) over the 3-year epoch difference in $C_{6,0}$, $C_{8,0}$, and $C_{10,0}$, we find that the differences between the two geopotential models are in fact reduced by a factor of 3 or more, suggesting an even closer level of agreement than simply difference the coefficients as published. \footnote{A minor point is also that the absolute value of the differences of $\bar{C}_{6,0}$, for example for GOCO05S and\\ JYY\textunderscore GOCE04S, and ITU\textunderscore GRACE16 and JYY\textunderscore GOCE04S, provided respectively with three and four significant digits in I2017, are erroneous. For example the value of $\bar{C}_{6,0}$ for model GOCO05S is evaluated at epoch January 1, 2008, neglecting annual variations. For model ITU\textunderscore GRACE16 the $\bar{C}_{6,0}$ value represents a mean for the period April 2009 to October 2013, and for model JYY\textunderscore GOCE04S a mean for the period November 2009 to October 2013: $\bar{C}_{60_{GOCO05S}} = -1.499663394539 \cdot 10^{-7}$, $\bar{C}_{60_{ITU\textunderscore GRACE16}} = -0.149998273044598 \cdot 10^{-6}$ and $\bar{C}_{60_{JYY\textunderscore GOCE04S}} = -0.1499850880263456 \cdot 10^{-6}$, so that their difference can be built as in I2017:
$\bar{C}_{60_{GOCO05S}} - \bar{C}_{60_{JYY\textunderscore GOCE04S}} =  1.87486 \cdot 10^{-11}$ and
$\bar{C}_{60_{ITU\textunderscore GRACE16}} - \bar{C}_{60_{JYY\textunderscore GOCE04S}}= 1.3185 \cdot 10^{-11} $.
However, in table (3) of I2017, these differences are respectively incorrectly quoted as $1.37 \cdot 10^{-11}$ and $1.827 \cdot 10^{-11}$, with four significant digits. These are not major errors but do influence the results to some degree.}

\subsubsection{\bfseries{Other inconsistent results in the publications by Iorio}}

It is curious that Iorio, in similar past papers \cite{ior1,ior2,ior3}, has produced results quite at variance with the present one in I2017, and with each other.
For example in 2005 \cite{ior1}, Iorio used the same technique that we applied to get a 5\% test of frame-dragging \cite{ciuetal16}  to predict a ``{\\itshape reliable}'' 1\% test of frame-dragging: ``{\itshape . . . by inserting the new spacecraft in a relatively low, and cheaper, orbit ($a$  = 7500 - 8000 km, $i$ _= 70 deg) and
suitably combining its node  with those of LAGEOS and LAGEOS II in order to
cancel out the first even zonal harmonic coefficients of the multipolar expansion of
the terrestrial gravitational potential $J_2, J_4$ along with their temporal variations.
The total systematic error due to the mismodelling in the remaining even zonal
harmonics would amount to $1\%$ and would be insensitive to departures of the
inclination from the originally proposed value of many degrees}'' \cite{ior1}.

But in a 2009 paper
 \cite{ior2} he claimed that the total measurement uncertainty of frame-dragging
including the LARES satellite, could range from $1000\%$  to $100\%$: ``{\itshape The low altitude of LARES, $1450 km$
with respect to about\\ $6000 km$ of LAGEOS and LAGEOS II,
will make its node sensitive to much more even zonals than its two already orbiting
twins; it turns out that, by using the sigmas of the covariance matrices of some of
the latest global Earth's gravity solutions based on long data sets of the dedicated
GRACE mission, the systematic bias due to the mismodeled even zonal harmonics up to $l  = 70$ will amount to $100 - 1000\%$}''  \cite{ior2}.
Later on, in 2011 \cite{ior3}, for the same orbit of the LARES satellite: ``{\itshape If, instead, one assumes $J_l$,
$l = 2,4,6,. . .$, i.e., the standard deviations of the sets of all the best estimates of $J_l$
for the models considered here the systematic bias, up to $l  = 60$, amounts to $12\%$
(SAV) [sum of absolute values] and $6\% (RSS)$ [root sum squared]. Again, also this
result may turn out to be optimistic for the same reasons as before.}''
Other similar papers published an uncertainty of 29\% for the LARES experiment \cite{ior4}.
Similar contradicting statements and huge differences for the uncertainty in the test of frame-dragging with the
LAGEOS and LAGEOS 2 satellites, published between 2003 and 2011, can be found
in other papers by the same author. In summary the author of I2017 has over about a decade published
error budgets of the same LARES experiment that go from 1000\% to 1\% with a number of figures in between.

\subsection{Conceptual shortcomings of differencing the lowest even zonals of different Earth gravity field models}

In I2017 the difference between the even zonals of different Earth gravity field models are calculated and then these differences are propagated into the nodal rates to find the total uncertainty in the measurement of frame-dragging.
However, as we remarked in a number of papers \cite{ciuetal10}, it makes $no$ sense to compare Earth gravity models obtained with different techniques that have different intrinsic accuracies (that is, including systematic errors and not simply formal errors) and especially that have different accuracies of the lowest harmonics. Indeed the accuracy of the lowest even zonal harmonics of an Earth gravity field model obtained with data of GOCE $only$, such as JYY\textunderscore GOCE04S, cannot be compared to the accuracy of the lowest harmonics of models obtained with GRACE and SLR. Furthermore, the accuracy of the lowest harmonics of a model obtained with an energy integral method, such as ITU\textunderscore GRACE16, should not be compared to that
of GGM05S; energy integral methods incorporate only instantaneous position determinations, without equations of motion to interpolate between subsequent measurements. For this reason, of the four Earth models (ITSG\textunderscore Grace2014S, GOCO05S, ITU\textunderscore GRACE16, JYY\textunderscore GOCE04S) used in I2017, only the lowest harmonics of ITSG\textunderscore Grace2014S and GOCO05s are comparable in accuracy to those of GGM05S. (We reiterate that I2017 does not carry out this comparison.)

Let us explain this point in detail. Satellite gravity gradiometry (SGG) is a very powerful technique for the direct observation of higher order functionals of the gravitational potential directly, rather than inferring them from their perturbing effects on satellite orbits. This is very nicely discussed in several articles, e.g., \cite{rum}. One of the drawbacks of SGG however is the fact that the observations are primarily sensitive to a range of frequencies of the geopotential, those that correspond to the measurement band of the specific instrument used. In the case of the GOCE mission, because of restrictions on the development of the gradiometer, the useful bandwidth was from $5 \cdot 10^{—3}Hz$ to $0.1 Hz$. In the end the very long wavelength components of the field cancel out in the measurement process as “common mode” effects that cannot rise over the noise of the instrument.

\begin{figure*}
\centering
 \includegraphics[width=0.9\textwidth]{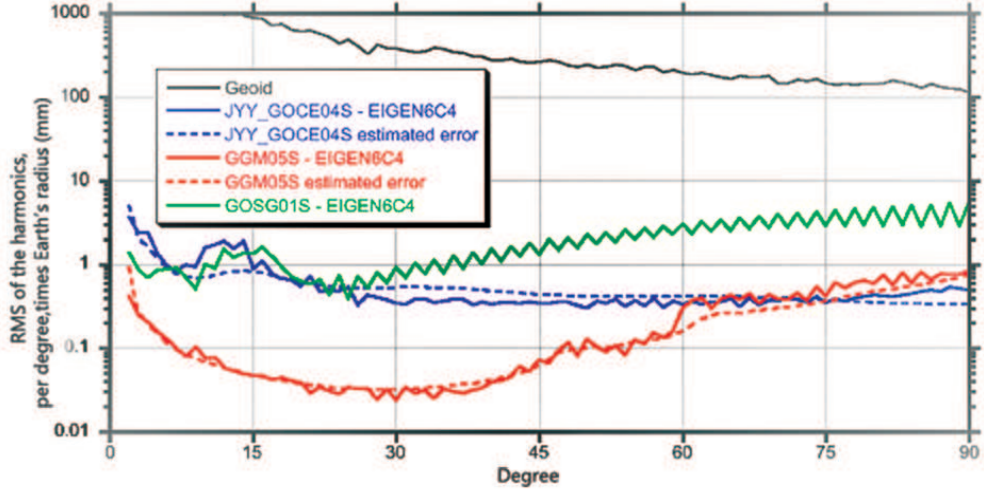}
\caption{We compare two GOCE-only gravity models (GOSG01S \cite{GOS} and JYY\textunderscore GOCE04S) as well as GGM05S with the EIGEN6C4 \cite{EIGEN6} (a ``combination model" which incorporates SST and SLR input to obtain highly accurate low-degree Earth gravity determinations - see text) \cite{reisEtAl2016,xuEtAl2017,yiEtAl2013,forsteEtAl2014}. The square-root variance (or RMS) is plotted as a function of the geopotential degree (the value of $2n$ in the symbol $J_{2n}$ of a multipole). The lower degrees represent longer  wavelength features of the gravity field. Included on the plot are the estimated errors assigned to JYY\textunderscore GOCE04S and to GGM05S, which appear to be consistent with the actual errors as realized by their differences with EIGEN6C4. At the higher degrees, the GOCE-based models perform slightly better than GRACE models, but for the purpose of the Lense-Thirring analysis, only the lowest degrees are relevant. It is clear that for the GOCE-only models, the lower degree terms are about an order of magnitude less accurate. They obviously perform even worse for degrees 10 to 16}
\label{fig:1}       
\end{figure*}

This results in SGG requiring some external information for the long wavelength (low degree) part of the field. This is the reason why those who use SGG data resort to adding-in Satellite-to-Satellite Tracking (SST) data, from which they can obtain the required information for the complete recovery of the field, from the lowest to the highest degree possible. In most cases the SST part comes from “high-low” Global Navigation Satellite System (GNSS) observations between the spacecraft carrying the gradiometer and GNSS spacecraft, and in such cases, the orbits are usually done in Precise Kinematic mode, which means that there are no equations of motion involved and the positions are determined independently at each observation point. This causes further degradation of the information contained in the very long wavelength part of these models. In other cases the information is derived from “low-low” SST, e.g. between the two GRACE spacecraft, using the ultra-precise K/Ka-Band Ranging (KBR) system, the same one used to produce the GRACE models. In that case of course the resulting model is a mix of GRACE and GOCE product, where the long wavelength info comes from the GRACE data and the higher degree part from GOCE, with the intermediate wavelengths being a region where both systems contribute.

Over the past two decades it was also recognized that due to the mass redistribution of the Earth System, the geopotential field is not a static one, it rather exhibits variations at all frequencies, spatial and temporal. Due to this, it is now customary that when one develops a model, these variations should be either estimated simultaneously, or forward-modeled on the basis of the best available models. For the longest wavelength components represented by the very low degree zonals, these are the estimates that we obtain from the analysis of several SLR missions covering several years, and these are part of the GRACE mission models. Obviously, models that are based on kinematic orbits (e.g. ITU\textunderscore GRACE, JYY\textunderscore GOCE) and use data over a short period of time, are not able to determine these temporal variations, but even worse, in most cases they do not even account for them, making it impossible to reference their coefficients to a specific date for comparison with models that are derived for a specific date (e.g. the GRACE mission models). Because of the high precision of the new techniques and the increase in modeling accuracy, temporal variations are now clearly visible up to high degrees and orders, so that comparison of models without careful consideration of these variations does not make any sense. GRACE has dealt with this issue by carefully developing a “de-aliasing” product that accounts for atmospheric, oceanic and such variations, so that the recovered variations can be ascribed to hydrological sources. Due to this specificity, it is no longer meaningful to use a single value and a linear rate to model even the very long wavelength components of the field (e.g. $J_2$). We now use a time-series of 15-day averaged values (sometimes even weekly estimates), in order to capture the effect of high frequency modulations caused by mass redistribution. One needs to be careful that these time series are derived using the same higher order model as in the case of the GRACE mission products, so that the ensemble represents the same potential field at all times (including the tidal part of course).

It is in the nature of the gravity gradiometer data from GOCE that the measurement errors dominate at the longer wavelength (lower degree) components of the gravity field. In Fig. 1, we compare two GOCE-only gravity models (GOSG01S and JYY\textunderscore GOCE04S) as well as GGM05S with the combination model EIGEN6C4. The square-root variance (or RMS) is plotted as a function of the geopotential degree (the value of $2n$ in the symbol $J_{2n}$ of a multipole). The lower degrees represent longer wavelengths  and higher degrees reflect the shorter wavelength features of the gravity field. Included on the plot are the estimated errors assigned to JYY\textunderscore GOCE04S and to GGM05S, which appear to be consistent with the actual errors as realized by their differences with EIGEN6C4. At the higher degrees, the GOCE-based models perform slightly better than GRACE-only models, but for the purpose of the Lense-Thirring analysis, only the lowest degrees are relevant. It is clear that for the GOCE-only models, the lower degree terms are about an order of magnitude less accurate and cannot rationally be used to judge the accuracy of gravity models that are based on GRACE data \footnote{All models mentioned are available at: http://icgem.gfz-potsdam.de/tom, along with the related documentation} \cite{reisEtAl2016,xuEtAl2017,yiEtAl2013,forsteEtAl2014}.

\begin{table*}[h!]
\centering
\begin{tabular}{p{7.5cm}|p{2.5cm}p{2.5cm}p{2.5cm}l}
    &  $\bar{C}_{6,0}$     &        $\bar{C}_{8,0}$        &   $\bar{C}_{10,0}$    \\\hline
Difference (absolute value) of GGM05S with ITSG\textunderscore Grace2014S & $5.72392 \cdot 10^{-13}$    & $9.35295 \cdot 10^{-13} $ &   $2.80392 \cdot 10^{-12}$  \\\hline
Difference (absolute value) of GGM05S with GOCO05S  & $8.84729 \cdot 10^{-12}$  &          $2.74188 \cdot 10^{-12}$    &      $2.28925 \cdot 10^{-12}$   \\

\end{tabular}
\caption{Difference of the even zonal harmonics $\bar{C}_{6,0}$, $\bar{C}_{8,0}$ and $\bar{C}_{10,0}$ between GGM05S, and ITSG\textunderscore Grace2014S and GOCO05S}
\end{table*}

\begin{table*}[h]
\centering
\begin{tabular}{p{7.5cm}|p{2.5cm}p{2.5cm}p{2.5cm}}
                                                                                              &  $\bar{C}_{6,0}$     &        $\bar{C}_{8,0}$        &   $\bar{C}_{10,0}$   \\\hline
Absolute value of the error propagated into the combination of the nodes of LAGEOS, LAGEOS 2 and LARES of the difference between GGM05S with ITSG\textunderscore Grace2014S in units of mas/yr &  0.0339082     & 0.00296451 &   0.258112  \\\hline

Absolute value of the error propagated into the combination of the nodes of LAGEOS, LAGEOS 2 and LARES of the difference between GGM05S with GOCO05S in units of mas/yr   & 0.524109   &          0.00869065    &      0.210734   \\

\end{tabular}
\caption{Error propagated into the node of LAGEOS, LAGEOS 2 and LARES due the differences between GGM05S and ITSG\textunderscore Grace2014S and GOCO05s for each coefficient $\bar{C}_{6,0}$, $\bar{C}_{8,0}$ and $\bar{C}_{10,0}$.}
\end{table*}

\begin{table*}[h]
\centering
\begin{tabular}{p{3.5cm}|p{10.5cm}}
                                                                                              & Total percent error relative to the  combined frame-dragging effect  \\\hline
ITSG\textunderscore Grace2014S 					& 0.588\% \\
GOCO05s							& 1.48\%   \\

\end{tabular}

\caption{Total error (sum of each absolute value) propagated into the combination of the nodes of LAGEOS, LAGEOS 2 and LARES relative to the combined frame-dragging effect of LAGEOS, LAGEOS 2 and LARES (about 50.465 mas/yr)}
\end{table*}

Naturally, the approach adopted in deriving a model and the amount of proper accounting of other-than-gravity variations of the ``observed''  field affect the accuracy of the derived model. The ``formal''  covariance that comes out as a product of a least squares estimation has very little to do with the true accuracy of the model. Calibrating this covariance matrix is usually the most time-consuming effort for most of the highest accuracy models and the developers make sure to report that process in detail when delivering their models. There are very few models that provide all the information required to judge them in a relative comparison to other models with similar information. Unfortunately, a blindly executed direct comparison ignoring all the details behind the development of two models, the reference epoch of the harmonic coefficients, the background models used, etc., most certainly leads to incorrect and unacceptable conclusions. Even models that are seemingly derived from similar data and using even the same technique, if they are based on data collected over two different time periods (even if of equal length), will be significantly different if the temporally varying parts are not appropriately handled in both cases. This reason alone ought to be enough to force a very strict approach in making comparisons between models.
A simple difference of the corresponding coefficients is definitely the wrong approach and especially one should not compare the lowest harmonics of ITSG\textunderscore Grace2014S and GOCO05S with those of ITU\textunderscore GRACE16 and JYY\textunderscore GOCE04S (this last gravity model being obtained with GOCE $only$), and then should not propagate these differences into the nodal rates to evaluate the uncertainty in the test of frame-dragging, as done in I2017.  ITSG\textunderscore Grace2014S and GOCO05S are models designed to be accurate for low order harmonics, so for completeness, in the next section we report the results of the errors obtained by differencing the lowest harmonics of ITSG\textunderscore Grace2014S and GOCO05S against the model GGM05S we use and then propagating these differences into the nodal rates. This approach  fully confirms our error budget in our test of frame-dragging. (To reiterate, I2017 did not consider comparisons to GGM05S.)

\subsection{Errors induced by the gravity field uncertainties}

We wish to compare the gravity field models\\ ITSG\textunderscore Grace2014S and GOCO05S with GGM05S. Therefore, we took the differences between each of the harmonics $\bar{C}_{6,0}$, $\bar{C}_{8,0}$ and $\bar{C}_{10,0}$ of GGM05S with the corresponding harmonics of the gravity field models
ITSG Grace2014S and GOCO05S (the differences are reported in Table 1). We then propagated these errors into the combination of the nodal rates, Table 2, and we finally added the absolute values of the errors due to each difference of each coefficient of these two gravity models and compared the result to the frame-dragging effect.
The results, shown in Table 3, obtained in this way, estimate the uncertainty in the GGM05S measurement of frame-dragging by modeling errors as (schematically) GMM05S - ITSG\textunderscore Grace2014S and GMM05S - GOCO05S.
The results shown in Table 3 are fully consistent with the systematic error budget of about 5\%, or less, for our test of frame-dragging \cite{ciuetal16}; in fact they are substantially smaller than that $5\%$ estimate.

\section{The erroneous unnecessary number of decimal digits of the coefficients $c_1$ and $c_2$ claimed to be necessary in I2017}

In I2017, it is claimed that ``the numerical values of $c_1$, $c_2$ in Eqs. (14), (15) are quoted with nine decimal digits in order to assure a cancelation of $J_2$ accurate to better than 1\% level.''

Let first explain why these coefficients are needed and how they are calculated. Our analysis is performed in the following way.

(1) We first obtain the residuals of the nodes of LAGEOS, LAGEOS 2 and LARES by using the experimental data, i.e., the Satellite Laser Ranging (SLR) observations of these satellites and by using, independently, the orbital estimators GEODYN (NASA), EPOS-OC (GFZ) and UTOPIA (CSR-UT). (The three estimators give consistent results.) The orbital residuals are the difference between the $observed$ orbital elements of a satellite, obtained by fitting the SLR observations using the three independent orbital estimators, and the $calculated$ orbital elements, obtained by propagating their orbits using the three orbital estimators containing a full set of physical models among which is an Earth gravity field model, such as GGM05S. The orbital residuals are mainly due by errors in the modelling of the orbital perturbations, such as errors in the spherical harmonic expansion of the Earth's gravity field, or to any perturbation not included at all in the orbital estimators, such as the Lense-Thirring effect. The main sources of error in the measurement of frame-dragging (see sections 1 and 2.1, and \cite{ciu96,ciuetal10}), which produce non-zero orbital residuals, are due to the lowest order even zonal harmonics of the Earth gravity field and in particular to the Earth quadrupole moment $C_{2,0}$ and to $C_{4,0}$.

(2) We then consider the system containing the three equations of the measured nodal residuals of
LAGEOS, LAGEOS 2 and LARES, $\delta \Omega$, in the three unknowns $\delta \bar{C}_{2,0}$, $\delta \bar{C}_{4,0}$ and Lense-Thirring
effect, parametrized by a parameter $\mu$, where $\mu$ is equal to unity in General Relativity. The three equations for LAGEOS, LAGEOS 2 and LARES are:

\begin{dmath}
\footnotesize
\delta \dot \Omega_{SAT}  \, =  \frac{3}{2} \, n_{SAT} \, \left ( \, \frac{R_{\oplus}}{a_{SAT}} \right )^2 \, \frac{cos \, I_{SAT}}{\left ( \, 1 - e_{SAT}^2 \,\right )^2 } \,\\
\Biggl\{ \,\sqrt{5} \delta \bar{C}_{20} + \sqrt{9} \delta \bar{C}_{40} \, \Biggl [ \, \frac{5} {8} \, \left ( \,
\frac{R_{\oplus}} {a_{SAT}} \, \right )^{2} \, \times ( \, 7 \, sin^2 \, I_{SAT} - 4 \, ) \,
\frac{( \, 1 + \frac{3}{2} \, e_{SAT}^2 )} {\left ( \, 1 - e_{SAT}^2 \, \right )^2} \, \Biggr ] +  \Sigma \, N_{2n \; SAT} \times \bar{C}_{2n \; 0} \, \Biggr \}+\mu \dot { \Omega\/}^{Lense-Thirring}_{SAT}
\end{dmath}

\noindent where SAT stands for LAGEOS or LAGEOS 2 or LARES, $n_{SAT}$ is their mean motion, $N_{2n \, SAT}$ are the coefficients (in the equation for the nodal rate) of the $\bar{C}_{2n, 0}$ for $2n > 4$, and the $\bar{C}_{2n,0}$ are  the normalized
even zonal harmonic coefficients, .

(3) We then solve for the frame-dragging effect, one of the three unknowns, together with $\delta \bar{C}_{20}$ and $\delta \bar{C}_{40}$,
and we get the frame-dragging effect as a function of the three residuals of the nodes of LAGEOS, LAGEOS 2 and LARES.
The result for frame-dragging, is:

\begin{equation}
\mu = \frac {\delta \Omega_{LAGEOS} + c_1 \delta \Omega_{LAGEOS \, 2} + c_2  \delta \Omega_{LARES}}
{{ \Omega}^{Lense-Thirring}_{LAGEOS} + c_1 { \Omega}^{Lense-Thirring}_{LAGEOS 2} +
c_2 { \Omega\/}^{Lense-Thirring}_{LARES}}
\end{equation}

Where the two coefficients $c_1$ and $c_2$ are $c_1 = 0.345$ and $c_2 = 0.073$.
The precise value of these two coefficients was not provided in \cite{ciuetal16} since they are updated every 15-arc as a function of the changes in the orbital parameters. Nevertheless, in \cite{ciupav} the values of these coefficients, in the case of the LAGEOS and LAGEOS 2 test of frame-dragging, were explicitly given.
Now I2017 provides in Eqs. 14 and 15 these coefficients with a large number of unnecessary decimal digits, claiming that at least nine significant decimal digits are needed for our test of frame-dragging.  However I2017 missed the main point of the technique that we used, as explained here and in a number of previous papers (see, e.g., \cite{ciupav,ciuetal10}. Indeed, the typical average size of the nodal residuals of the LAGEOS and LAGEOS 2, using the most recent determinations of the Earth gravity field, is of the order of about 150 mas/yr. Since the frame-dragging effect has on LAGEOS and LAGEOS 2 a size of about 31 mas/year, for a 5\% measurement of frame-dragging, thus with an error of about $\pm$ 1.5 mas/yr, the coefficient $c_1$ of LAGEOS 2, must only be accurate, at the level of about 1\%, i.e., two significant decimal digits of the $c_1$ are enough for a 5\% test, similarly two/three significant decimal digits of the LARES coefficient $c_2$ are enough for a 5\% test. Thus, contrary to what is claimed in I2017 the two coefficients $c_1$ and $c_2$ are only needed at the level of two or three significant decimal digits. I2017 misunderstood the analysis technique, and missed also this basic point.
Nevertheless we determined these two coefficients with many more significant digits, thanks to the technique of SLR to measure all the orbital elements of LAGEOS, LAGEOS 2 and LARES.

\section{Brief review of the methods to combine the orbital elements and results by other groups confirming our test}

The use of two passive laser-ranged satellites of LAGEOS type, with supplementary inclinations, to test\\ frame-dragging
was proposed in \cite{ciu84,ciu86,ciu89,csr89,asi89,rie89,pet,ciu96}.
The combination of the nodes of a number of satellites, used in \cite{ciuetal16} , was first proposed in \cite{ciu89} (see page 3102).
Then in \cite{ciu96} was first calculated the precise combination of the orbital elements of LAGEOS and LAGEOS 2. In
\cite{ciupav}  the combination of the nodes of LAGEOS and LAGEOS 2 was displayed and used to provide a
test of frame-dragging. In \cite{ciu06} the use of the nodes of LAGEOS and LAGEOS 2 and of a similar satellite at
a lower altitude (LARES) was proposed; the uncertainty in the measurement of frame-dragging using these three satellites was
then calculated as a function of the inclination and of the semimajor axis of LARES (see Fig. 2). These calculations coupled with the
capabilities of the first qualifying launcher VEGA, led to the precise orbit of the LARES successfully launched
in 2012 by VEGA.

\begin{figure}
 \includegraphics[width=0.9\textwidth]{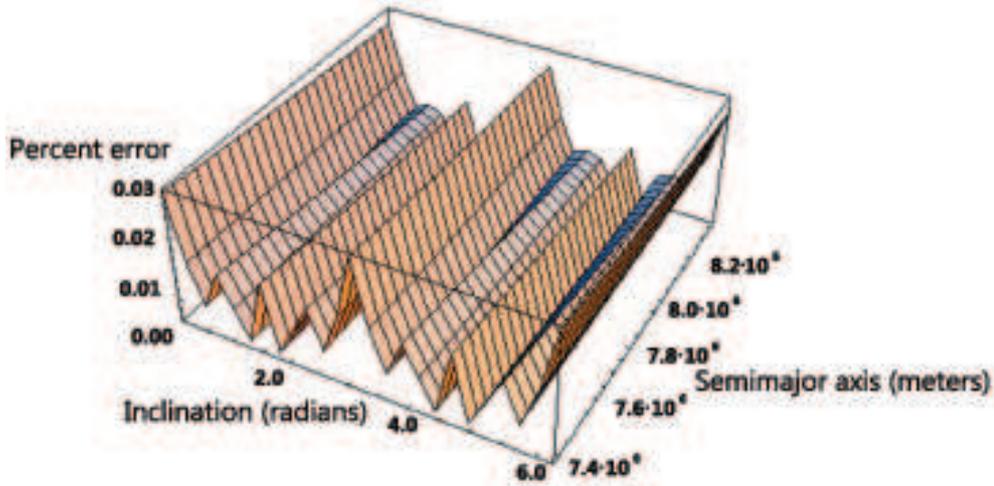}
\caption{Percent error in the measurement of the Lense-Thirring effect, due
to the even zonal harmonics uncertainties, as a function of the inclination
and of the semimajor axis of LARES, using LARES, LAGEOS and LAGEOS 2.
The range of the semimajor axis of LARES is between 7400 km and 8300 km
and that of the inclination between 0 and 2 $\Pi$ [adapted from \cite{ciu06}.}
\label{fig:2}       
\end{figure}

Furthermore in I2017 it is claimed ``Finally, it is remarkable that, after about twenty years since
the first reported tests with LAGEOS and LAGEOS II and
four years since the launch of LARES, nobody has yet published
any genuinely independent test of the Lense–Thirring
effect with such geodetic satellites in the peer-reviewed literature,
especially in view of how many researchers around the
world constitute the global satellite laser ranging community.''
I2017 seems to be unaware of the fact that the three
$independent$ orbital estimators GEODYN, EPOS-OC and UTOPIA have been
$independently$ run, respectively by the three groups of: (a) Universities of Salento (Lecce), Sapienza
(Rome), and Maryland BC/JCET (Joint Center for Earth Systems Technology);
(b) Center for Space Research (CSR) of the University
of Texas (UT) at Austin \cite{rie08,rie09} and GFZ (German Research Centre for Geosciences, Helmholtz Centre, Potsdam) \cite{koe12,koe16},
leading to the same results.
Furthermore, the test published in 2016 in \cite{ciuetal16} was fully confirmed by another completely
independent team and presented at an international conference \cite{bas}.
A similar test of frame-dragging, the 19\% test of frame-dragging by Gravity Probe B \cite{GPB}, was indeed
published by one team only.

\section{Conclusions}

All the claims of I2017 are groundless. They are either numerically and conceptually incorrect
or are based on erroneous assumptions and claims. In section (2.1) we have shown that the numerical figures
of I2017 are erroneous by some large factor; in section (2.2) we have explained that the lowest harmonics of
different Earth gravity field models, e.g., those obtained with GOCE only, such as JYY\textunderscore GOCE04S, and those obtained with GRACE and SLR,
such as GGM05S, cannot be compared and thus I2017 is flawed by the incorrect assumption of comparing the
lowest harmonics of different,  noncomparable, Earth gravity models.
We also reported that by comparison of low degree harmonics of suitable, comparable, gravity field models, the 5\% systematic error estimate
of our Lense-Thirring analysis is confirmed.
In section 3 we showed that it is incorrect to
claim that the coefficients used in the combination of the satellites’ residual nodal rates
must be known with nine significant decimal digits, indeed three significant decimal
digits are enough for a 1\% test of frame-dragging.
Finally, in section 4, we evidenced that the\\ LARES test of frame-dragging was indeed
repeated by independent and different teams, contrary to the claims in I2017.
\section*{Acknowledgments}
We gratefully acknowledge the Italian Space Agency for the support of the LARES and LARES 2 space missions
under agreements No. 2017-23-H.0 and No. 2015-021-R.O. We are also grateful to the International Ranging Service,
ESA, AVIO and ELV. ECP acknowledges the support of NASA Grants NNX09AU86G and NNX14AN50G. RM
acknowledges NASA Grant NNX09AU86G and NSF Grant PHY-1620610 and JCR the support of NASA Contract
NNG17V105C. We thank the anonymous referee for useful comments to improve the paper.

\end{document}